\documentclass[12pt,a4paper]{article}
\usepackage{amssymb}
\usepackage{amsfonts}
\usepackage[english]{babel}
\usepackage{graphicx}

\newcommand{\bean}{\begin{eqnarray*}}
\newcommand{\eean}{\end{eqnarray*}}
\newcommand{\ba}{\begin{array}}
\newcommand{\ea}{\end{array}}
\newcommand{\pa}{\partial}

\newcommand{\no}{\nonumber}

\newcommand{\be}{\begin{equation}}
\newcommand{\ee}{\end{equation}}
\newcommand{\bea}{\begin{eqnarray}}
\newcommand{\eea}{\end{eqnarray}}
\newcommand{\beaa}{\begin{eqnarray*}}
\newcommand{\eeaa}{\end{eqnarray*}}

\begin{document}

\title
{The Interactions of Solitons in the Novikov-Veselov Equation}
\author{
 Jen-Hsu Chang \\Department of Computer Science and Information Engineering, \\
 National Defense University, \\
 Tauyuan County 33551, Taiwan }

\date{}

\maketitle
\begin{abstract}
Using the reality condition of the solutions, one constructs the
real Pfaffian N-solitons solutions of the Novikov-Veselov (NV)
equation using the $\tan$ function and the Schur identity. By the
minor-summation formula of the Pfaffian, we can study the
interactions  of solitons in the Novikov-Veselov equation from the
Kadomtsev-Petviashvili (KP) equation's  point of view, that is, the
totally non-negative  Grassmannian. Especially, the Y-type
resonance, O-type and the P-type interactions of X-shape are
investigated. Also, the maximum amplitude of the intersection of the
line solitons and the critical angle are  computed and one makes a
comparison with the KP-(II)  equation.
\end{abstract}

\newpage

\section{Introduction}
\subsection{Novikov-Veselov Equation }
The Novikov-Veselov equation \cite{ba,gm,nv,vn}  is defined by
($U$ and $t$ is real) \bea
 U_t &=& \pa_z^3 U+ {\bar{\pa}}_z^3 U+ 3\pa_z(VU)+3 {\bar{\pa}}_z
 (\bar{V}U ), \label{NV} \\
  {\bar{\pa}}_z V&=&\pa_z U. \no
 \eea
 When $z=\bar{z}=x$, we get the famous KdV equation ($U=V=\bar{V}$)
 \[ U_t=2U_{xxx}+12UU_x. \]
  The equation (\ref{NV}) can be represented as the form of Manakov's triad \cite{ma}
 \[H_t=[A, H]+BH, \]
 where $H$ is the two-dimension Schrodinger operator
 \[H= \pa_z \bar{\pa_z} +U\]
 and
 \[A= \pa_z^3 + V\pa_z+ {\bar{\pa}}_z^3+\bar{V}{\bar{\pa}}_z,
 \quad B= V_z+\bar{V}_{\bar{z}}.\]
 It is equivalent to the linear representation
 \bea H \phi=0, \quad \pa_t \phi=A \phi. \label{rep} \eea
 We see that the Novikov-Veselov equation (\ref{NV}) preserves a
 class of the purely potential self-adjoint operators $H$. Here
 the pure potential means $H$ has no external electric and
 magnetic fields. The periodic inverse spectral problem for the
 two-dimensional Schrodinger operator $H$ was investigated in
 terms of the Riemann surfaces with some group of involutions and
 the
 corresponding Prym $\Theta$-functions \cite{dk, gn, kr, no,sh}. On the other hand, it is  known that the Novikov-Veselov
hierarchy is a
 special reduction of the two-component BKP hierarchy \cite{wz,kt}(and references
 therein). In \cite{wz}, the authors showed that the Drinfeld-Sokolov
 hierarchy of D-type is a reduction of the  two-component BKP
 hierarchy using two different types of pseudo-differential
 operators, which is different from Shiota's point of view \cite{sh}. Finally, it is worthwhile to notice that the Novikov-Veselov
 equation (\ref{NV})
 is a special reduction of the Davey-Stewartson equation
  \cite{ko, kg}.
 \\ \indent Let $H \phi= H \omega=0.$ Then via the Moutard transformation
 \cite{an, mo, ni}
 \beaa U(z, \bar{z}) &\longrightarrow & \hat{U}(z, \bar{z})=U(z, \bar{z})
 +2\pa \bar{\pa} \ln \omega \\
 \phi &\longrightarrow & \theta=\frac{i}{\omega} \int (\phi\pa \omega-\omega\pa \phi)dz-(\phi\bar{\pa} \omega-\omega\bar{\pa}
 \phi)d \bar{z},
 \eeaa
one can construct a new Schrodinger operator $\hat{H}= \pa_z
\bar{\pa_z} +\hat{U}$ and  $\hat {H} \theta=0.$   \\
\indent The extended Moutard transformation was established such
that $\hat{U}(t,z,\bar{z})$ and $\hat{V}(t,z,\bar{z})$ defined by
\cite{hh, ms} \beaa \hat{U}(t,z,\bar{z}) &=& U(t,z,\bar{z})+2\pa
\bar{\pa} \ln  W(\phi, \omega)  \\
\hat{V} (t,z,\bar{z}) &=& V(t,z,\bar{z})+2 \pa \pa \ln  W (\phi,
\omega), \eeaa where the skew product (alternating bilinear form)
$W$ is defined by
 \bea W (\phi, \omega) &=& \int (\phi\pa
\omega-\omega\pa \phi)dz-(\phi\bar{\pa} \omega-\omega\bar{\pa}
 \phi)d \bar{z}+[\phi\pa^3 \omega-\omega\pa^3
 \phi+\omega\bar{\pa}^3 -\phi\bar{\pa}^3
 \omega   \no \\
&+&
2(\pa^2\phi\pa\omega-\pa\phi\pa^2\omega)-2(\bar{\pa}^2\phi\bar{\pa}\omega-\bar{\pa}\phi\bar{\pa}^2\omega)
 +3 V(\phi\pa \omega-\omega\pa \phi) \no \\
 &-&3\bar{V}(\phi\bar{\pa} \omega
 -\omega\bar{\pa}
 \phi)]dt , \label{ext}
 \eea
will also satisfy the Novikov-Veselov equation. \\
\indent Next, we construct Pfaffian-type solutions. Given any $N$
wave functions  $\phi_1, \phi_2, \phi_3, \\
\cdots, \phi_N $ (or their linear combinations) of (\ref{rep}) for
fixed potential $U(z, \bar{z}, t)$ , the N-step extended Moutard
transformation can be obtained in the Pfaffian \cite{an, ni} (
also see \cite{he, oh})
\[P(\phi_1, \phi_2, \phi_3, \cdots, \phi_N)=\left
\{\begin{array}{ll} Pf(\phi_1, \phi_2, \phi_3, \cdots, \phi_N), &
N \quad \mbox{even} , \\
\widetilde{Pf}(\phi_1, \phi_2, \phi_3, \cdots, \phi_N), & N \quad
\mbox{odd} ,\end{array} \right. \]
 \bea Pf(\phi_1, \phi_2,
\phi_3, \cdots, \phi_N) &=&
\sum_{\sigma}\epsilon(\sigma)W_{\sigma_1 \sigma_2}W_{\sigma_3
\sigma_4} \cdots W_{\sigma_{N-1} \sigma_N} \label{pf1} \\
\widetilde{Pf}(\phi_1, \phi_2, \phi_3, \cdots, \phi_N) &=&
\sum_{\sigma}\epsilon(\sigma)W_{\sigma_1 \sigma_2}W_{\sigma_3
\sigma_4} \cdots W_{\sigma_{N-2} \sigma_{N-1}}\phi_{\sigma_N},
\label{pf2} \eea
 where $W_{\sigma_{i} \sigma_j}
=W(\phi_{\sigma(i)}, \phi_{\sigma(j)})$ is defined  by the skew
product (\ref{ext}). \noindent The summations $\sigma$ in
(\ref{pf1}) and (\ref{pf2}) run from over the permutations of
$\{1,2,3,\cdots, N \}$ such that $ \sigma_1 < \sigma_2, \sigma_3 <
\sigma_4, \sigma_5 < \sigma_6, \cdots $ and
\[ \sigma_1 < \sigma_3 < \sigma_5 <\sigma_7 \cdots ,\]
with $\epsilon(\sigma)=1$ for the even permutations and
$\epsilon(\sigma)=-1$ for the odd permutations. Then the solution
$U$ and $V$ can be expressed as \cite{an} \bea U &=& U_0+ 2 \pa
\bar{\pa}
[ \ln P(\phi_1, \phi_2, \phi_3, \cdots, \phi_N)] \no \\
V &=& V_0+ 2 \pa \pa [ \ln P(\phi_1, \phi_2, \phi_3, \cdots,
\phi_N)]. \label{pf3} \eea
\indent Recently, the resonance theory of line solitons of KP-(II) equation
\[   \pa_x (-4 u_t+u_{xxx}+6uu_x)+ 3u_{yy}=0  \]
has attracted much attractions using the totally non-negative Grassmannian \cite{ad, bc, ch, ko, ko1, ko3, ko4}, that is, those points of the real Grassmannian whose Plucker coordinates are all non-negative.  For the KP-(II) equation case, the  $\tau$-functiom  is described by the Wroskian form with respect to $x$ . Inspired by their works, one can consider the Novikov-Veselov equation similarly. In this article, we study the basic interactions of real solitons, i.e., the Y-type resonance, O-type and P-type solitons of X-shape of  the NV equation from the totally non-negative Grassmannian.  One soliton solution is described in detail. The maximum amplitude of the intersection of X-shape  and critical angle are computed and one makes a comparison with the KP-(II) equation. It is shown that the maximum amplitude of X-shape has a  different behavior from KP-(II), that is, they depend on the extra parameter  $\epsilon$ (see below); however, one thinks that the resonance theory of solitons in the NV equation has parallel structure with the KP-(II) equation. \\
\indent The paper is organized as follows. In section 2, one constructs the real soliton solution using the $\tan$ function and the Schur Identity. By the minor-summation formula of the Pfaffian, we can study the interactions  of solitons in the Novikov-Veselov equation from the totally non-negative  Grassmannian. Furthermore,  one soliton solution is described in detail. In section 2, we investigate the basic interactions of line solitons and the maximum amplitude of the intersection of X-shape  and critical angle are computed  and one makes a comparison with the KP-(II) equation.
In section 4,  we conclude the paper with several remarks.

\section{Real N-Solitons Potentials }
In this section, one introduces the real Grassmannian (or the $ 2N \times M$ matrix) to construct N solitons and explains  the conditions for real potentials. Also,  one soliton solution is described in detail. \\
\indent To obtain the N-solitons solutions, we assume that $V=0$
in (\ref{NV}) and recall that $\pa \bar{\pa}=\frac{1}{4} \triangle
$.
One considers $U=-\epsilon \neq 0$, i.e.,  \bea  \pa \bar{\pa} \phi &=& \epsilon \phi \no \\
\phi_t &=& \phi_{zzz}+\phi_{\bar{z}\bar{z}\bar{z}},
\label{pri} \eea where $\epsilon$ is non-zero real constant.  The
general solution of (\ref{pri}) can be expressed as \be \phi(z,
\bar{z}, t) = \int_{\Gamma} e^{(i\lambda) z+(i\lambda)^3
t+\frac{\epsilon}{i\lambda}\bar{z}+\frac{\epsilon^3}{(i\lambda)^3}t
} \nu(\lambda) d \lambda, \label{con} \ee where $\nu(\lambda)$ is
an arbitrary distribution and $\Gamma$ is an arbitrary path of
integration such that the RHS of (\ref{con}) is well defined. \\
\indent Next, using (\ref{pf1}) and (\ref{con}), one can construct
the N-solitons solutions. Let's take $\nu_m(\lambda)=\delta
(\lambda-p_m)$, where $p_m$ is complex numbers. Then one defines \be
\phi_m =  \frac{\phi (p_m)}{\sqrt{3}}= \frac{1}{\sqrt{3}}e^{F(p_m)}, \label {fun} \ee where
\[ F(\lambda)=(i\lambda)z+(i\lambda)^3t+\frac{\epsilon}{i\lambda}\bar{z}+\frac{\epsilon^3}{(i\lambda)^3}t. \]
Then a direct calculation of the extended Moutard
transformation (\ref{ext}) can yield
\be W(\phi_m,\phi_n) = i \frac{p_n-p_m}{p_n+p_m} e^{F(p_m)+F(p_n)}. \label{bac} \ee
To introduce the real Grasssmannian (or the resonance), we have to consider linear combination of $\phi_n$.
Let's assume that \[\vec{\phi}=(\phi_1,
\phi_2, \phi_3, \cdots, \phi_M)^T
\] and $H$ be an $2N \times M ( 2N \leq M)$ of real constant  matrix (or Grassmannian). We
suppose
\[ H \vec{\phi}=\vec{\phi}^\ast=(\phi_1^\ast, \phi_2^\ast, \phi_3^\ast, \cdots,
\phi_{2N}^\ast)^T,  \]
that is,
\[\phi_n^\ast= h_{n1} \phi_1+ h_{n2} \phi_2+h_{n3} \phi_3+ \cdots +h_{nM} \phi_M, \quad 1\leq n \leq 2N.\]
Then it can be shown that
\begin{equation} Pf( \phi_1^\ast, \phi_2^\ast, \phi_3^\ast, \cdots, \phi_{2N}^\ast )=Pf (H W_M H^T), \end{equation}
where the $M \times M$ matrix $W_M$ is defined by the element (\ref{bac}).
Now, we have the
minor-summation formula \cite{is, km}
\begin{equation} \tau_N =Pf (H W_M H^T)=\sum_{I
\subset [M],\quad \sharp I=2N} Pf(H_I^I)det (H_I),
\label{sum}
\end{equation}
 where
$H_I^I$ denote the $2N \times M $ submatrix of $H$ obtained by
picking up the rows and columns indexed by  the same index set
$I$. From this formula, one can investigate the possibility of
resonance of real solitons of the Novikov-Veselov equation using
the resonance theory of KP-(II) equation \cite{ko1, ko2, ko3}. Finally,
the N-solitons solutions are
defined by \cite{bd, jh1}
\be U(z, \bar{z}, t)=- \epsilon + 2 \pa
\bar{\pa} \ln \tau_N (z, \bar{z}, t), \quad V(z, \bar{z}, t)=2 \pa
\pa \ln \tau_N (z, \bar{z}, t). \label{solu} \ee

Given $n+m=N$ pairs of complex numbers
\bean
&& (p_1, q_1), ( p_2, q_2), \cdots,  ( p_n, q_n) \\
&& ( \lambda_1, \mu_1), ( \lambda_2, \mu_2), \cdots(\lambda_m, \mu_m), \eean to get the real potential $U$, we have
the following reality conditions \cite{bd}:
\begin{itemize}
\item Type (I): \\
\be   p_{\ell} \bar{q}_{\ell}=-
\epsilon, \quad {\ell}=1,2 ,3 , \cdots, n    \label{ty} \ee
\item Type (II): \\ \bean  |\lambda_k|^2= |\mu_k|^2= \epsilon > 0, \quad k=1,2,3, \cdots, m \eean
\item Type (III): Mixed type of Type (I) and Type (II)
\end{itemize}
 \indent We remark that when $\epsilon \to \pm \infty $, the
Veselov-Novikov equation reduces to the KP-I ( $\epsilon \to - \infty $) and KP-(II) ($\epsilon \to \infty$ ) equation
respectively \cite{gri}. To make a comparison with KP-(II) equation, we consider only the resonance of type
(II). Then we have, letting $ p_m= \sqrt{\epsilon} e^{ i \alpha_m}
$ and {\bf removing $i$ factor  from (\ref{bac}) afterwards} , \[
W(\phi_m,\phi_n)= -\tan\frac{\alpha_n-\alpha_m}{2}
e^{\phi_{mn}}, \] where
 \bea \phi_{mn}&=&
F(p_m)+F(p_n)=-2\sqrt{\epsilon}[x(\sin\alpha_m+
\sin\alpha_n)+y(\cos\alpha_m+ \cos \alpha_n)]  \no \\
&+& 2t\epsilon \sqrt{\epsilon}(\sin3\alpha_m+\sin3\alpha_n) \label{phi}
\eea
Then we have
\bea \tau_1 &=& \tan \frac{\alpha_1-\alpha_2}{2} e^{\phi_{12}}+ a \tan \frac{\alpha_1-\alpha_3}{2} e^{\phi_{13}} \\
&=& a e^{\phi_{13}} \tan \frac{\alpha_1-\alpha_3}{2} [ 1+ \frac{1}{a} \frac{\tan \frac{\alpha_1-\alpha_2}{2}}{\tan \frac{\alpha_1-\alpha_3}{2}} e^{F(p_2)-F(p_3)}]  \\
&=&  a e^{\phi_{13}} \tan \frac{\alpha_1-\alpha_3}{2} [ 1+  e^{F(p_2)-F(p_3)+\theta_{23}}] , \\
\eea
where $a$ is a constant and the phase shift
\[ \theta_{23}=\ln \frac{1}{a} \frac{\tan \frac{\alpha_1-\alpha_2}{2}}{\tan \frac{\alpha_1-\alpha_3}{2}}=\ln \frac{\tan \frac{\alpha_1-\alpha_2}{2}}{\tan \frac{\alpha_1-\alpha_3}{2}}-\ln a \]
Hence the real one-soliton solution is
\bea
U &=& -\epsilon+ 2 \pa_{z}\pa_{\bar{z}} \ln a e^{\phi_{13}} \tan \frac{\alpha_1-\alpha_3}{2} [ 1+  e^{F(p_2)-F(p_3)+\theta_{23}}] \\
&=& -\epsilon+ 2 \pa_{z}\pa_{\bar{z}} [ 1+ e^{F(p_2)-F(p_3)+\theta_{23}}] \\
&=&  -\epsilon+ \frac{1}{2} |p_3-p_2|^2 sech^2 [\frac{F(p_2)-F(p_3)+\theta_{23}}{2}] \\
&=& -\epsilon + 2 \epsilon \sin^2(\frac{\alpha_3-\alpha_2}{2})sech^2[\frac{F(p_2)-F(p_3)+\theta_{23}}{2}]  \\
&=& -\epsilon + A_{[2,3]} sech^2 \frac{1}{2} \left(\vec{\bf{K}}_{[2,3]} \cdot \vec{x}-{\bf{\Omega}}_{[2,3]}t+\theta_{23} \right). \eea
From (\ref{phi}) the amplitude $ A_{[2,3]} $ , the wave vector $\vec{\bf{K}}_{[2,3]}$ and the frequency ${\bf\Omega}_{[2,3]}$ are defined by
\bea
A_{[2,3]} &=& 2 \epsilon \sin^2(\frac{\alpha_3-\alpha_2}{2}) \\
\vec{\bf{K}}_{[2,3]} &=& 2 \sqrt{\epsilon} (-\sin \alpha_2+\sin \alpha_3, -\cos \alpha_2+ \cos \alpha_3) \label{k}\\
{\bf\Omega}_{[2,3]}  &=& 2 \epsilon \sqrt{\epsilon} [-\sin 3 \alpha_2+\sin 3\alpha_3] \label{o}
\eea
The direction of the wave vector $\vec{\bf{K}}_{[2,3]}= (K_{[2,3]}^x, K_{[2,3]}^y)$ is measured in the clockwise sense from the y-axis and it is given by
\be \frac{K_{[2,3]}^y}{K_{[2,3]}^x}= \frac{-\cos \alpha_2+ \cos \alpha_3}{-\sin \alpha_2+\sin \alpha_3}=-\tan\frac{\alpha_2+ \alpha_3}{2}, \label{ang} \ee
that is, $\frac{\alpha_2+ \alpha_3}{2}$ gives the angle between the line $\vec{\bf{K}}_{[2,3]} \cdot \vec{x}$=constant and the y-axis.\\
\indent Furthermore, the wave vector $\vec{\bf{K}}_{[2,3]}$ and the frequency $\bf{\Omega}_{[2,3]}$ satisfies the soliton-dispersion relation after a simple calculation
\[{\bf{\Omega}}_{[2,3]}=\epsilon K_{[2,3]}^x\left(\frac{3(K_{[2,3]}^y)^2- (K_{[2,3]}^x)^2)}{4 \epsilon} +  \frac{3(K_{[2,3]}^x)^2-9(K_{[2,3]}^y)^2}{ (K_{[2,3]}^x)^2+(K_{[2,3]}^y)^2}            \right). \]

The soliton velocity $\bf{V_{[2,3]}}$ is along the direction of the wave vector $\vec{\bf{K}}_{[2,3]}$, and is defined by $ \vec{\bf{K}}_{[2,3]}\cdot \bf{V_{[2,3]}}= \bf{\Omega}_{[2,3]}       $, which yields
\[  {\bf{V_{[2,3]}}}=\frac{\bf{\Omega}_{[2,3]}}{\vert \vec{\bf{K}}_{[2,3]}\vert^2}  \vec{\bf{K}}_{[2,3]}=\frac{\epsilon}{4} \frac{\sin 3 \alpha_3-\sin 3 \alpha_2}{\sin^2\frac{\alpha_2-\alpha_3}{2}} \left( \sin\alpha_3-\sin \alpha_2,  \cos  \alpha_3-\cos \alpha_2\right)                                                  \]
Also,
\beaa
 \Lambda_2 (z, \bar{z}, t) &=& Pf (W_4)= Pf
 \left[\ba{cccc} W(p_1, p_1) & W(p_1, p_2) & W(p_1, p_3)  &  W(p_1, p_4)    \\
W(p_2, p_1) & W(p_2, p_2)  & W(p_2, p_3) & W(p_2, p_4)  \\
W(p_3, p_1)  & W(p_3, p_2) & W(p_3, p_3) & W(p_3, p_4)  \\
 W(p_4, p_1)  & W(p_4, p_2) & W(p_4, p_3) &  W(p_4, p_4)  \\
\ea \right] \\
 &=& Pf
 \left[\ba{cccc} 0 & -\tan\frac{\alpha_2-\alpha_1}{2}
e^{\phi_{12}} & -\tan\frac{\alpha_3-\alpha_1}{2} e^{\phi_{13}}  &
-\tan\frac{\alpha_4-\alpha1}{2}
e^{\phi_{14}}    \\
\tan\frac{\alpha_2-\alpha_1}{2} e^{\phi_{21}} & 0 &
-\tan\frac{\alpha_3-\alpha_2}{2} e^{\phi_{23}}&
-\tan\frac{\alpha_4-\alpha_2}{2}e^{\phi_{24}}  \\
\tan\frac{\alpha_3-\alpha_1}{2} e^{\phi_{31}}  &
\tan\frac{\alpha_3-\alpha_2}{2} e^{\phi_{32}} & 0 &
-\tan\frac{\alpha_4-\alpha_3}{2}
e^{\phi_{43}}  \\
\tan\frac{\alpha_4-\alpha_1}{2}e^{\phi_{41}} &
\tan\frac{\alpha_4-\alpha_2}{2} e^{\phi_{42}} &
\tan\frac{\alpha_4-\alpha_3}{2}
e^{\phi_{43}} &  0  \\
\ea \right] \\
&=& \tan\frac{\alpha_2-\alpha_1}{2}
\tan\frac{\alpha_3-\alpha_1}{2} \tan\frac{\alpha_4-\alpha_1}{2}
\tan\frac{\alpha_3-\alpha_2}{2}
\tan\frac{\alpha_4-\alpha_2}{2}\tan\frac{\alpha_4-\alpha_3}{2} \\
&\times & e^{F(p_1)+F(p_2)+F(p_3)+F(p_4)} \eeaa
Here we have utilized the formula that if $\mathbf{C}$ is a
$2N \times 2N$  matrix  with $(i,j)$-th entry $
\frac{\alpha_i-\alpha_j}{\alpha_i+\alpha_j}$, then one has the
Schur identity  \cite{nim,os} \be Pf(\mathbf{C})=\prod_{1 \leq i <
j \leq 2N} ( \frac{\alpha_i-\alpha_j}{\alpha_i+\alpha_j}).
\label{ca} \ee
Similarly, \beaa
 && \Lambda_3 (z, \bar{z}, t) =Pf(W_6) \\
  &=& Pf
\left[\ba{cccccc} W(p_1, p_1) & W(p_1, p_2) & W(p_1, p_3)  & W(p_1, p_4)  & W(p_1, p_5) & W(p_1, p_6)  \\
W(p_2, p_1) & W(p_2, p_2) & W(p_2, p_3)  & W(p_2, p_4)  & W(p_2, p_5) & W(p_2, p_6)  \\
W(p_3, p_1) & W(p_3, p_2) & W(p_3, p_3)  & W(p_3, p_4)  & W(p_3, p_5) & W(p_3, p_6)  \\
W(p_4, p_1) & W(p_4, p_2) & W(p_4, p_3)  & W(p_4, p_4)  & W(p_4, p_5) & W(p_4, p_6)  \\
W(p_5, p_1) & W(p_5, p_2) & W(p_5, p_3)  & W(p_5, p_4)  & W(p_5, p_5) & W(p_5, p_6)  \\
W(p_6, p_1) & W(p_6, p_2) & W(p_6, p_3)  & W(p_6, p_4)  &
W(p_6,p_5) & W(p_6, p_6)  \\ \ea \right] \\
&=& -[\tan\frac{\alpha_2-\alpha_1}{2}
\tan\frac{\alpha_3-\alpha_1}{2}
\tan\frac{\alpha_4-\alpha_1}{2}\tan\frac{\alpha_5-\alpha_1}{2}\tan\frac{\alpha_6-\alpha_1}{2}
\tan\frac{\alpha_3-\alpha_2}{2}  \\
& \times & \tan\frac{\alpha_4-\alpha_2}{2}
\tan\frac{\alpha_5-\alpha_2}{2}\tan\frac{\alpha_6-\alpha_2}{2}\tan\frac{\alpha_4-\alpha_3}{2}
\tan\frac{\alpha_5-\alpha_3}{2} \tan\frac{\alpha_6-\alpha_3}{2} \\
& \times  & \tan\frac{\alpha_5-\alpha_4}{2}
\tan\frac{\alpha_6-\alpha_4}{2} \tan\frac{\alpha_6-\alpha_5}{2}]
\times  e^{F(p_1)+F(p_2)+F(p_3)+F(p_4)+F(p_5)+F(p_6)}\eeaa
From
the $\Lambda_2$ and $\Lambda_3$, we require the order relation of
$\alpha_i $ to study the resonance of N-solitons:
\[  -\frac{\pi}{2} < \alpha_1< \alpha_2 <\alpha_3 < \cdots < \alpha_{2N-1} < \alpha_{2N} < \frac{\pi}{2}.         \]
Actually, it is not difficult to see from (\ref{ca}) that
\be \Lambda_N(x,y,t)= Pf (W_{2N}) =(-1)^N \left (\prod_{i=2,\, i> j}^{2N} \tan \frac{\alpha_i-\alpha_j}{2}\right) e^{\sum_{m=1}^{2N}  F(p_m)}. \label{la} \ee
Therefore, given a $ 2N \times M $ matrix $H$, the associated $\tau_H$-function can be written as by (\ref{sum}) and (\ref{la})
\be \tau_H= \sum_{I \subset [M],\quad \sharp I=2N} \Upsilon_I \Lambda_I(x,y,t), \label{h} \ee
where $ \Upsilon_I $ is the $2N \times 2N $ minor for the columns with the index set  $I={ i_1, i_2, i_3, \cdots, i_{2N}} $ and
\[   -\frac{\pi}{2} < \alpha_1< \alpha_2 <\alpha_3 < \cdots < \alpha_{M-1} < \alpha_{M} < \frac{\pi}{2}.    \]
{\bf Remark:} The relation of $\tau_H$ with  the $\tau$-function \cite{bc} of the KP-(II) equation.\\
We see that
\be F(p_n)=-2\sqrt{\epsilon}[x \sin\alpha_n+y\cos \alpha_n]+ 2t\epsilon \sqrt{\epsilon}\sin3\alpha_n.  \label{T}        \ee
Now, we consider the double scaling limit: let $\epsilon \to \infty$ and $ \alpha_n \to 0$ such that
\be k_n= -2\sqrt{\epsilon} \alpha_n \quad is \quad fixed .\label{app} \ee
Using the formula as $\alpha_n \to 0$
\[ \sin \alpha_n \approx \alpha_n- \frac{\alpha_n^3}{3!}, \quad \cos \alpha_n \approx 1- \frac{\alpha_n^2}{2!} ,\]
one has
\[  F(p_n) \approx x k_n+ Y k_n^2 + T k_n^3, \]
where
\[ y= 4 \sqrt{\epsilon} Y, t= \frac{8}{9} T.\]
Furthermore, one sees that
\[ \pa_y= \frac{1}{ 4 \sqrt{\epsilon}} \pa_Y \]
and then
\[ \pa_{xx}+ \pa_{yy} \approx \pa_{xx}  \]
as $\epsilon \to \infty$.
Finally, since
\[  \tan \frac{\alpha_m-\alpha_n}{2} \approx \frac{\alpha_m-\alpha_n}{2}= \frac{1}{ 2 \sqrt{\epsilon}}(k_n-k_m), \]
we have under the condition (\ref{app})
\[   \tau_H= \tau_{KP},  \]
up to  some of little significant overall multiplicative factor.

\section{Interactions of Solitons}
In this section, we investigate three basic interactions of solitons, i.e. Y-type (resonance), O-type and P-type.
\begin{itemize}
\item Y-type interaction:
We consider the matrix
\[ H_Y=\left[\ba{ccc} 1 & 0 & -b    \\ 0 & 1 & a   \ea
 \right]. \]
 where $a,b$ are positive number.  By the formula (\ref{sum}), the corresponding $\tau$-function is
 \[ \tau_Y= \tan\frac{\alpha_1-\alpha_2}{2}e^{\phi_{12}}+a \tan\frac{\alpha_1-\alpha_3}{2}e^{\phi_{13}} +b \tan\frac{\alpha_2-\alpha_3}{2}e^{\phi_{23}}. \]
Using (\ref{phi}) and (\ref{sum}), one has  (see figure 1)\\
(a) For $y >> 0$, there are two unbounded line solitons, whose types from left to right are
\[  [1,2], [2,3] \]
(b) For $ y << 0$, there is one  unbounded line soliton, whose type is
\[  [3,1]. \]
Also, the corresponding permutation is (231). From this, one can conjecture that the rank condition in the  KP-(II) equation is also correct for $y >> 0$ and $ y << 0$ from (\ref{ang}) in the Novikov-Veselov equation; for example \cite{ko4} , considering the permutation (671823945), we have \\
(a) For $y >> 0$, there are four  unbounded line solitons, whose types from left to right are
\be   [1,6], [2,7],[4,8], [7,9] \label{o1} \ee
(b) For $y << 0$, there is five  unbounded line soliton, whose types from left to right are.
\be  [9,5],[8,4], [6,3], [5,2], [3,1]. \label{o2} \ee
\hspace*{0.3cm} On the other hand, from (\ref{k}) and (\ref{o}), it can be seen that one has the following resonant conditions for wave number and frequency
\[ \vec{\bf{K}}_{[1,3]}= \vec{\bf{K}}_{[1,2]}+ \vec{\bf{K}}_{[2,3]}, \quad   {\bf\Omega}_{[1,3]} ={\bf\Omega}_{[1,2]} + {\bf\Omega}_{[2,3]}. \]

\item O-type soliton \\
 In this case, we consider the matrix
 \[ H_O=\left[\ba{cccc} 1 & a & 0 & 0  \\ 0 & 0 & 1 & b  \ea
 \right], \]
where  $a, b$ are positive numbers. Then
\[ H_O [\phi(p_1), \phi(p_2), \phi(p_3), \phi(p_4)]^T =\left[\ba{c} \phi(p_1)+a \phi(p_2)
  \\   \phi(p_3)+ b\phi(p_4)   \ea \right]= \left[\ba{c}
  \mu_1  \\  \mu_2  \ea \right]. \]
A direct calculation yields by (\ref{sum})
\begin{equation}
\tau_O=W(\mu_1,\mu_2)= [W(p_1, p_3) + b W(p_1, p_4) + a W(p_2, p_3)+ ab W(p_2, p_4)]. \label{oo}
\end{equation}

For the $[1,2]$-soliton, one knows that  by (\ref{solu})  (see figure 2) \\
(a)When $ y >> 0 $ (the upper left), $ \tau_O \approx W(p_1, p_4)+ a W(p_2, p_4)  \Rightarrow $
\[ U \approx  -\epsilon + 2 \epsilon \sin^2(\frac{\alpha_2-\alpha_1}{2})sech^2[\frac{F(p_1)-F(p_2)+\theta_{12}^{+}}{2}], \]
where
\[ \theta_{12}^{+}= \ln \tan\frac{\frac{\alpha_3-\alpha_1}{2}}{\frac{\alpha_3-\alpha_2}{2}}-\ln a .\]
(b)When  $ y << 0 $ (the bottom right), $\tau_O \approx b W(p_1, p_3)+ ab W(p_2, p_3)  \Rightarrow $
\[ U \approx  -\epsilon + 2 \epsilon \sin^2(\frac{\alpha_2-\alpha_1}{2})sech^2[\frac{F(p_1)-F(p_2)+\theta_{12}^{-}}{2}], \]
where
\[ \theta_{12}^{-}= \ln \tan\frac{\frac{\alpha_4-\alpha_1}{2}}{\frac{\alpha_4-\alpha_2}{2}}-\ln a .\]
The total shift is
\[ \theta_{12}= \theta_{12}^{+}-\theta_{12}^{-}= \ln\frac{ \tan\frac{\alpha_3-\alpha_1}{2} \tan \frac{\alpha_4-\alpha_2}{2}}{ \tan  \frac{\alpha_3-\alpha_2}{2} \tan   \frac{\alpha_4-\alpha_1}{2}}  \]
Similarly, for the $[3,4]$-soliton, one knows that \\
(a)When $ y >> 0 $ (the upper right), $\tau_O \approx W(p_1, p_3)+ b W(p_1, p_4)  \Rightarrow $
\[ U \approx  -\epsilon + 2 \epsilon \sin^2(\frac{\alpha_4-\alpha_3}{2})sech^2[\frac{F(p_3)-F(p_4)+\theta_{34}^{+}}{2}], \]
where
\[ \theta_{34}^{+}= \ln \tan\frac{\frac{\alpha_3-\alpha_1}{2}}{\frac{\alpha_4-\alpha_1}{2}}-\ln b .\]
(b)When  $ y << 0 $ (the bottom left), $\tau_O \approx a W(p_2, p_3)+ ab W(p_2, p_4)  \Rightarrow $
\[ U \approx  -\epsilon + 2 \epsilon \sin^2(\frac{\alpha_4-\alpha_3}{2})sech^2[\frac{F(p_3)-F(p_4)+\theta_{34}^{-}}{2}], \]
where
\[ \theta_{34}^{-}= \ln \tan\frac{\frac{\alpha_3-\alpha_2}{2}}{\frac{\alpha_4-\alpha_2}{2}}-\ln b .\]
The total shift is
\[ \theta_{34}= \theta_{34}^{+}-\theta_{34}^{-}= \ln\frac{ \tan\frac{\alpha_3-\alpha_1}{2} \tan \frac{\alpha_4-\alpha_2}{2}}{ \tan  \frac{\alpha_3-\alpha_2}{2} \tan   \frac{\alpha_4-\alpha_1}{2}}=\theta_{12}=\ln \Delta_O,   \]
where
\[ \Delta_O = \frac{ \tan\frac{\alpha_3-\alpha_1}{2} \tan \frac{\alpha_4-\alpha_2}{2}}{ \tan  \frac{\alpha_3-\alpha_2}{2}\tan \frac{\alpha_4-\alpha_1}{2}} .\]

From $\Delta_O$, one can consider the periodic function
\[  f(x)=     \frac{\tan\frac{(x-\alpha_2)}{2}}{\tan \frac{(x-\alpha_1)}{2}}, \quad    \frac{\pi}{2} \geq x \geq  \alpha_2 > \alpha_1 \geq \frac{-\pi}{2}.            \]
It can be seen that $\Delta_O > 1 $ in the increasing interval and $0< \Delta_O < 1 $ in the decreasing interval. A direct calculation shows that critical point of $f(x)$ is
\[ x =\frac{\alpha_1+\alpha_2+\pi}{2}.\]
Hence the function of $f(x)$ has a critical point if $\alpha_1+\alpha_2 < 0$ and it is always increasing if $\alpha_1+\alpha_2 \geq 0$ \\
 \hspace*{0.3cm} Also, each $[i,j]$-soliton shifts in $x$ with
\[ \Delta x_{[i,j]}=\ \frac{1}{\sin \alpha_j-\sin\alpha_i} \theta_{ij}.  \]

\hspace*{0.3cm} Next, as in the KP-(II) case \cite{ko}, we compute the amplitude of the intersection. We place the soliton solution so that the origin (0,0) is the center of the X-shape at $t=0$. This implies that the sum of the phase shifts becomes zero for each soliton, i.e.
\beaa
\theta_{12}^{+} +\theta_{12}^{-}&=& \ln \frac{\tan\frac{\alpha_3-\alpha_1}{2}\tan\frac{\alpha_4-\alpha_1}{2}}{\tan \frac{\alpha_3-\alpha_2}{2}\tan \frac{\alpha_4-\alpha_2}{2}}-2\ln a=0,  \\
\theta_{34}^{+} +\theta_{34}^{-}&=&  \ln \frac{\tan\frac{\alpha_3-\alpha_1}{2}\tan\frac{\alpha_3-\alpha_2}{2}}{\tan \frac{\alpha_4-\alpha_1}{2}\tan \frac{\alpha_4-\alpha_2}{2}}                                  -2\ln b=0.
\eeaa
These determine $a$ and $b$ in the matrix $A_O$. From $(\ref{oo})$, one has
\beaa
\tau_O &=& \tan \frac{\alpha_3-\alpha_1}{2}e^{F(p_1)+F(p_3)}+ \sqrt{\frac{\tan\frac{\alpha_3-\alpha_1}{2}\tan\frac{\alpha_3-\alpha_2}{2}}{\tan \frac{\alpha_4-\alpha_1}{2}\tan \frac{\alpha_4-\alpha_2}{2}} } \times \tan \frac{\alpha_4-\alpha_1}{2} \times e^{F(p_1)+F(p_4)} \\
&+& \sqrt{ \frac{\tan\frac{\alpha_3-\alpha_1}{2}\tan\frac{\alpha_4-\alpha_1}{2}}{\tan \frac{\alpha_3-\alpha_2}{2}\tan \frac{\alpha_4-\alpha_2}{2}}                                                         } \times \tan \frac{\alpha_3-\alpha_2}{2} \times e^{F(p_2)+F(p_3)} \\
&+& \frac{\tan \frac{\alpha_3-\alpha_1}{2}}{\tan \frac{\alpha_4-\alpha_2}{2}} \times \tan \frac{\alpha_4-\alpha_2}{2} \times e^{ F(p_2)+F(p_4)} \\
&=& \tan \frac{\alpha_3-\alpha_1}{2}[ e^{ F(p_1)+F(p_3)}+e^{ F(p_2)+F(p_4)}+ \sqrt{\frac{1}{\Delta_O}}( e^{ F(p_1)+F(p_4)}+e^{ F(p_2)+F(p_3)})] \\
&\equiv & e^{ F(p_1)+F(p_3)}+ \sqrt{\frac{1}{\Delta_O}} e^{ F(p_1)+F(p_4)}+ e^{ F(p_2)+F(p_4)}+\sqrt{\frac{1}{\Delta_O}} e^{ F(p_2)+F(p_3)} \\
& \equiv & \cosh \Theta_O^+ + \sqrt{\frac{1}{\Delta_O}}\cosh \Theta_O^- \equiv \sqrt{\Delta_O} \cosh \Theta_O^+ + \cosh \Theta_O^-,
\eeaa
where $\equiv$ means the same solution for $U$ and
\be  \Theta_O^{\pm} =\frac{1}{2} [(F(p_1)-F(p_2))\pm (F(p_3)-F(p_4))]. \label{to}\ee
Then after a simple calculation, we get from (\ref{solu}), denoting
\[ \hat U=U+\epsilon, \]
\beaa
 2 \hat U &=& (\pa_{xx} + \pa_{yy}) \ln \tau_O  \\
 &=& \frac{\sqrt{\Delta_O}\cosh \Theta_O^+ \cosh \Theta_O^-[(\Theta_{Ox}^+)^2+  (\Theta_{Oy}^+)^2 +  (\Theta_{Ox}^-)^2 +(\Theta_{Oy}^-)^2]}{[\sqrt{\Delta_O}\cosh \Theta_O^+ + \cosh \Theta_O^-]^2} \\
 &+& \frac{\Delta_O [(\Theta_{Ox}^+)^2+(\Theta_{Oy}^+)^2 ] -2\sqrt{\Delta_O} \sinh \Theta_O^+ \sinh \Theta_O^- [\Theta_{Ox}^+ \Theta_{Ox}^- + \Theta_{Oy}^+ \Theta_{Oy}^- ] }
 {[\sqrt{\Delta_O}\cosh \Theta_O^+ + \cosh \Theta_O^-]^2},
 \eeaa
where by (\ref{to})
\beaa
\Theta_{Ox}^{\pm} &=& -\sqrt{\epsilon}[\sin\alpha_1-\sin\alpha_2 \pm (\sin\alpha_3-\sin\alpha_4)] \\
\Theta_{Oy}^{\pm} &=& -\sqrt{\epsilon}[\cos\alpha_1-\cos\alpha_2 \pm (\cos\alpha_3-\cos\alpha_4)].
\eeaa
At $\cosh  \Theta_O^+ = \cosh \Theta_O^-=1 $ (or $ \sinh  \Theta_O^+ = \sinh \Theta_O^-=0 $ ), we have the maximum of $\hat U$:
\be
2 \hat U^O_{max} =\frac{\sqrt{\Delta_O}-1}{\sqrt{\Delta_O}+1} [(\Theta_{Ox}^{+})^2+(\Theta_{Oy}^{+})^2]+ \frac{(\Theta_{Ox}^{+})^2+(\Theta_{Ox}^{-})^2+(\Theta_{Oy}^{+})^2+(\Theta_{Oy}^{-})^2}{\sqrt{\Delta_O}+1}. \label{max}
\ee
A direct calculation yields
\beaa
&& (\Theta_{Ox}^{+})^2+(\Theta_{Oy}^{+})^2=2[A_{[1,2]}+ A_{[3,4]} + 2 \sqrt{A_{[1,2]} A_{[3,4]}} \cos \frac{(\alpha_1+\alpha_2)-(\alpha_3+\alpha_4)}{2}] \\
&& (\Theta_{Ox}^{+})^2+(\Theta_{Ox}^{-})^2+(\Theta_{Oy}^{+})^2+(\Theta_{Oy}^{-})^2=4(A_{[1,2]}+ A_{[3,4]}),
\eeaa
where
\[ A_{[1,2]}= 2 \epsilon \sin^2 \frac{\alpha_1-\alpha_2}{2}, \quad A_{[3,4]}= 2 \epsilon \sin^2 \frac{\alpha_3-\alpha_4}{2}. \]

Hence one has
\bea
2 \hat U^O_{max} &=& \frac{\sqrt{\Delta_O}-1}{\sqrt{\Delta_O}+1}\{2[A_{[1,2]}+ A_{[3,4]} + 2 \sqrt{A_{[1,2]} A_{[3,4]}}\cos \frac{(\alpha_1+\alpha_2)-(\alpha_3+\alpha_4)}{2}]\} \no \\
 &+& \frac{4(A_1+A_2)}{\sqrt{\Delta_O}+1} \no \\
&=& 2 ( A_{[1,2]}+ A_{[3,4]}  ) + 4
\frac{\sqrt{\Delta_O}-1}{\sqrt{\Delta_O}+1} \sqrt{A_{[1,2]}
A_{[3,4]}}\cos \frac{(\alpha_1+\alpha_2)-(\alpha_3+\alpha_4)}{2}
\label{oma}  \eea

Then we see that \\
%\begin{itemize}
 $  \bullet  \quad 0 < \Delta_O < 1 : \quad (\sqrt{A_{[1,2]}}- \sqrt{A_{[3,4]}})^2 < \hat U^O_{max} <  ( A_{[1,2]}+ A_{[3,4]}) $   \\
 $  \bullet  \quad  \Delta_O >1  : \quad  ( A_{[1,2]}+ A_{[3,4]}  ) < \hat  U^O_{max} < (\sqrt{A_{[1,2]}}+ \sqrt{A_{[3,4]}})^2.
 $  \\
%\end{itemize}
For the case with $\alpha_1=\alpha_2$ (or $\alpha_3=\alpha_4$), one has $\Delta_O=1$. Then one of the line-soliton vanishes, and the limit consists of just one-soliton solution. For the case $\alpha_2=\alpha_3$, i.e., $\Delta_O=\infty$, the $\tau_O$ has only three terms, which corresponds to Y-type solution (i.e., the phase shift $\ln \Delta_O $ becomes infinity and the middle portion of the interaction stretches to infinity). It can be seen that  $ \Delta_O >1 $ will correspond to the KP-(II) case \cite{ko}. \\
\hspace*{0.3cm} Now, let's consider the special case when both solitons are of equal amplitude and symmetric with respect to y-axis, i.e.,
\[A_{[1,2]}= A_{[3,4]}=A, \quad \alpha_1=-\alpha_4 < 0, \quad \alpha_2=-\alpha_3 < 0. \]
One notices that the line-soliton  $A_{[3,4]}$ gives the angle  $\frac{\alpha_3+ \alpha_4}{2}$ between the line and y-axis measured in the clockwise sense from the y-axis and the line-soliton  $A_{[1,2]}$ gives the angle  $\frac{\alpha_1+ \alpha_2}{2}$ between the line and y-axis measured in the counter-clockwise sense from the y-axis. Then
\[ \tan\frac{\alpha_3+ \alpha_4}{2}=\tan(\frac{\alpha_4- \alpha_3}{2}+\alpha_3)\geq \tan \frac{\alpha_4- \alpha_3}{2}=\sqrt{\frac{A}{2\epsilon-A}},\]
which means for fixed amplitude $A$ and $\epsilon$ such that  $2\epsilon$  $>$  $ A $, the angle $\frac{\alpha_3+ \alpha_4}{2}(=-\frac{\alpha_1+ \alpha_2}{2})$ has a lower bound given by the critical angle, i.e., $\alpha_3=0$,
\[  \phi_C^O=\arctan\sqrt{\frac{A}{2\epsilon-A}}=\arcsin\sqrt{\frac{A}{2\epsilon}}.  \]
Similarly, one can introduce the following Miles-parameter \cite{ko, mi} to describe the interaction for the O-type solution
\[ \kappa=\frac{\tan \frac{\alpha_4+\alpha_3}{2}}{ \tan  \phi_C^O} = \frac{\tan \frac{\alpha_4+\alpha_3}{2}}{ \tan \frac{\alpha_4- \alpha_3}{2} } \geq 1.          \]
Then \[\Delta_O= \frac{\kappa^2-\delta^4}{\kappa^2-1}, \] where
\[ \delta= \tan \frac{\alpha_4+\alpha_3}{2}=\kappa \tan \phi_C^O.   \]
From (\ref{oma}), one has
\[\hat U^O_{max} = \frac{4A}{1+ \sqrt{\Delta_O}}
\frac{\sqrt{\Delta_O}+\delta^2}{1+\delta^2}.\]
Hence  at the critical angle, i.e., $\kappa=1$, one gets $ \hat  U^O_{max}=4A$ and the phase shift $\theta_{34}=\infty$, leading to the $Y$-shape interaction.

\item P-type soliton \\
 In this case, we consider the matrix
 \[ H_P=\left[\ba{cccc} 1 & 0 & 0 & -c  \\ 0 & 1 & a & 0  \ea
 \right], \]
where  $a, c$ are positive numbers. Then
\[ H_P [\phi(p_1), \phi(p_2), \phi(p_3), \phi(p_4)]^T =\left[ \ba{c} \phi(p_1)- c \phi(p_4)\\
\phi(p_2)+ a\phi(p_3)   \ea \right]= \left[\ba{c} \nu_1  \\  \nu_2  \ea \right]. \]
A direct calculation yields by (\ref{sum}) 
\[ \tau_P=W(\nu_1,\nu_2)= W(p_1, p_2) + a W(p_1, p_3) + c W(p_2, p_4)+ ac W(p_3, p_4). \]
For the $[2,3]$-soliton, one knows that  by (\ref{solu}) ( see figure 3) \\
(a)When $ y >> 0 $ (the upper right), $ \tau_P \approx W(p_1, p_2)+ a W(p_1, p_3)  \Rightarrow $
\[ U \approx  -\epsilon + 2 \epsilon \sin^2(\frac{\alpha_2-\alpha_3}{2})sech^2[\frac{F(p_2)-F(p_3)+\theta_{23}^{+}}{2}], \]
where
\[ \theta_{23}^{+}= \ln \tan\frac{\frac{\alpha_2-\alpha_1}{2}}{\frac{\alpha_3-\alpha_1}{2}}-\ln a .\]
(b)When  $ y << 0 $ (the bottom left) , $ \tau_P \approx c W(p_2, p_4)+ ac W(p_3, p_4)  \Rightarrow $
\[ U \approx  -\epsilon + 2 \epsilon \sin^2(\frac{\alpha_2-\alpha_3}{2})sech^2[\frac{F(p_2)-F(p_3)+\theta_{23}^{-}}{2}], \]
where
\[ \theta_{23}^{-}= \ln \tan\frac{\frac{\alpha_2-\alpha_4}{2}}{\frac{\alpha_3-\alpha_4}{2}}-\ln a .\]
The total shift is
\[ \theta_{23}= \theta_{23}^{+}-\theta_{23}^{-}= \ln\frac{ \tan\frac{\alpha_2-\alpha_1}{2} \tan \frac{\alpha_3-\alpha_4}{2}}{ \tan  \frac{\alpha_3-\alpha_1}{2}  \tan   \frac{\alpha_2-\alpha_4}{2}}  \]
Also, for the $[1,4]$-soliton, it is seen that  that \\
(a)When $ y >> 0 $ (the upper left), $\tau_P \approx W(p_1, p_3)+ c W(p_3, p_4)  \Rightarrow $
\[ U \approx  -\epsilon + 2 \epsilon \sin^2(\frac{\alpha_1-\alpha_4}{2})sech^2[\frac{F(p_1)-F(p_4)+\theta_{14}^{+}}{2}], \]
where
\[ \theta_{14}^{+}= \ln \tan\frac{\frac{\alpha_1-\alpha_2}{2}}{\frac{\alpha_2-\alpha_4}{2}}-\ln c .\]
(b)When  $ y << 0 $ (the bottom right), $\tau_P \approx a W(p_1, p_2)+ ac W(p_2, p_4)  \Rightarrow $
\[ U \approx  -\epsilon + 2 \epsilon \sin^2(\frac{\alpha_1-\alpha_4}{2})sech^2[\frac{F(p_1)-F(p_4)+\theta_{14}^{-}}{2}], \]
where
\[ \theta_{14}^{-}= \ln \tan\frac{\frac{\alpha_3-\alpha_2}{2}}{\frac{\alpha_4-\alpha_2}{2}}-\ln b .\]
The total shift is
\[ \theta_{14}= \theta_{14}^{+}-\theta_{14}^{-}= \ln\frac{ \tan\frac{\alpha_2-\alpha_1}{2} \tan \frac{\alpha_3-\alpha_4}{2}}{ \tan  \frac{\alpha_3-\alpha_1}{2} \tan   \frac{\alpha_2-\alpha_4}{2}}=\theta_{23}=\ln \Delta_P,   \]
where
\be \Delta_P = \frac{ \tan\frac{\alpha_2-\alpha_1}{2} \tan \frac{\alpha_3-\alpha_4}{2}}{ \tan \frac{\alpha_3-\alpha_1}{2}\tan \frac{\alpha_2-\alpha_4}{2}} .\label{P}\ee
From $\Delta_P$, one can similarly consider the periodic function
\[  g(x)=     \frac{\tan\frac{(x-\alpha_4)}{2}}{\tan \frac{(x-\alpha_1)}{2}}, \quad    \frac{\pi}{2}  \geq  \alpha_4 > x > \alpha_1 \geq \frac{-\pi}{2} .            \]
It is not difficult to show that the derivative  $g'(x)$ is positive for $x \in (\alpha_1, \alpha_4).$  Hence one knows that $0 < \Delta_P < 1$. \\
\hspace*{0.3cm} Similarly, each $[i,j]$-soliton shifts in $x$ with
\[ \Delta x_{[i,j]}=\ \frac{1}{\sin \alpha_j-\sin\alpha_i} \theta_{ij}.  \]
\hspace*{0.3cm} Now, one considers the amplitude of the
intersection of $P$-type.Then \beaa \tau_P =  \sqrt{\Delta_P} \cosh
\Theta_P^+ + \cosh \Theta_P^-, \eeaa and \be \Theta_P^{\pm}
=\frac{1}{2} [(F(p_1)-F(p_4))\pm (F(p_2)-F(p_3))]. \label{po}\ee
Then  a simple calculation yields \beaa
 2 \hat  U &=& (\pa_{xx} + \pa_{yy}) \ln \tau_P  \\
 &=& \frac{\sqrt{\Delta_P}\cosh \Theta_P^+ \cosh \Theta_P^-[(\Theta_{Px}^+)^2+  (\Theta_{Py}^+)^2 +  (\Theta_{Px}^-)^2 +(\Theta_{Py}^-)^2]}{[\sqrt{\Delta_P}\cosh \Theta_P^+ + \cosh \Theta_P^-]^2} \\
 &+& \frac{\Delta_P [(\Theta_{Px}^+)^2+(\Theta_{Py}^+)^2 ] -2\sqrt{\Delta_P} \sinh \Theta_P^+ \sinh \Theta_P^- [\Theta_{Px}^+ \Theta_{Px}^- + \Theta_{Py}^+ \Theta_{Py}^- ] }
 {[\sqrt{\Delta_P}\cosh \Theta_P^+ + \cosh \Theta_P^-]^2},
 \eeaa
where by (\ref{po}) \beaa
\Theta_{Px}^{\pm} &=& -\sqrt{\epsilon}[\sin\alpha_1-\sin\alpha_4 \pm (\sin\alpha_2-\sin\alpha_3)] \\
\Theta_{Py}^{\pm} &=& -\sqrt{\epsilon}[\cos\alpha_1-\cos\alpha_4 \pm
(\cos\alpha_2-\cos\alpha_3)]. \eeaa At $\cosh  \Theta_P^+ = \cosh
\Theta_P^-=1 $ (or $ \sinh  \Theta_P^+ = \sinh \Theta_P^-=0 $ ), we
have the maximum of $\hat U$: \be 2 \hat  U^P_{max}
=\frac{\sqrt{\Delta_P}-1}{\sqrt{\Delta_P}+1}
[(\Theta_{Px}^{+})^2+(\Theta_{Py}^{+})^2]+
\frac{(\Theta_{Px}^{+})^2+(\Theta_{Px}^{-})^2+(\Theta_{Py}^{+})^2+(\Theta_{Py}^{-})^2}{\sqrt{\Delta_P}+1}.
\label{maxp} \ee A direct  calculation yields \beaa
&& (\Theta_{Px}^{+})^2+(\Theta_{Py}^{+})^2=2[A_{[1,4]}+ A_{[2,3]} + 2 \sqrt{A_{[1,4]} A_{[2,3]}} \cos \frac{(\alpha_1+\alpha_4)-(\alpha_2+\alpha_3)}{2}] \\
&& (\Theta_{Px}^{+})^2+(\Theta_{Px}^{-})^2+(\Theta_{Py}^{+})^2+(\Theta_{Py}^{-})^2=4(A_{[1,4]}+
A_{[2,3]}), \eeaa where
\[ A_{[1,4]}= 2 \epsilon \sin^2 \frac{\alpha_1-\alpha_4}{2}, \quad A_{[2,3]}= 2 \epsilon \sin^2 \frac{\alpha_2-\alpha_3}{2}. \]
Finally,  we get \bea
2 \hat U^P_{max} &=& \frac{\sqrt{\Delta_P}-1}{\sqrt{\Delta_P}+1}\{2[A_{[1,4]}+ A_{[2,3]} + 2 \sqrt{A_{[1,4]} A_{[2,3]}}\cos \frac{(\alpha_1+\alpha_4)-(\alpha_2+\alpha_3)}{2}]\} \no \\
 &+& \frac{4(A_{[1,4]}+ A_{[2,3]})}{\sqrt{\Delta_P}+1} \no \\
&=& 2 ( A_{[1,4]}+ A_{[2,3]}  ) + 4
\frac{\sqrt{\Delta_P}-1}{\sqrt{\Delta_P}+1} \sqrt{A_{[1,4]}
A_{[2,3]}}\cos \frac{(\alpha_1+\alpha_4)-(\alpha_2+\alpha_3)}{2}
\label{pma}
\eea
\hspace*{0.3cm} Let's assume $\alpha_1+\alpha_4=0$, i.e., the $A_{[1,4]}$-soliton is the y-axis \cite{ch}. Then the angle between  $A_{[1,4]}$-soliton and $A_{[2,3]}$-soliton is $\frac{\alpha_2+\alpha_3}{2}=\rho$ in the clockwise sense when it is positive. We see that
\bea \sin |\rho| &=& \sin | \frac{\alpha_2+\alpha_3} {2}| <  \sin \frac{\alpha_2-\alpha_1}{2}  \leq \sin \frac{(\alpha_4-\alpha_3)+(\alpha_2-\alpha_1)}{2}  \no \\
& = & \sin \frac{(\alpha_4-\alpha_1)-(\alpha_3-\alpha_2)}{2}. \no
\eea
Hence for fixed amplitudes $A_{[1,4]}$, $A_{[2,3]}$ and $\epsilon$ such that  $2\epsilon$  $>$  $ A_{[1,4]}   >   A_{[2,3]} $, the angle $\rho$ between the two line solitons satisfies
\be -\phi_C^P < \rho < \phi_C^P, \quad \phi_C^P = \arcsin \sqrt{\frac{A_{[1,4]}}{2\epsilon}}- \arcsin \sqrt{\frac{A_{[2,3]}}{2\epsilon}}. \label{rho} \ee
To estimate the $U^P_{max}$, we have to compute, noting that $0 <\Delta_P < 1 $,
\[0<  C=\frac{1-\sqrt{\Delta_P}}{1+\sqrt{\Delta_P}} < 1 .\]
Now, we notice that
\beaa
\tan \frac{\alpha_2-\alpha_1}{2} &=&  \tan \frac{\phi_C^P+\rho}{2}, \quad  \tan \frac{\alpha_4-\alpha_3}{2} = \tan \frac{\phi_C^P-\rho}{2}  \\
\tan \frac{\alpha_3-\alpha_1}{2} &=&  \tan \frac{(\alpha_3-\alpha_2)+(\alpha_2-\alpha_1)}{2} \\
\tan \frac{\alpha_4-\alpha_2}{2} &=&  \tan \frac{(\alpha_4-\alpha_3)+(\alpha_3-\alpha_2)}{2}
\eeaa
Then by (\ref{P})
\beaa
\Delta_P &=& \left( 1-\frac{\sec^2 \frac{ \alpha_2-\alpha_1}{2}}{1+[\tan \frac{\alpha_2-\alpha_1}{2}/ \tan \frac{\alpha_3-\alpha_2}{2} ]} \right) \left( 1-\frac{\sec^2 \frac{ \alpha_4-\alpha_3}{2}}{1+[\tan \frac{\alpha_4-\alpha_3}{2}/ \tan \frac{\alpha_3-\alpha_2}{2} ]} \right)  \\
&=& \left( 1-\frac{2\sec (\phi_C^P+\rho)}{1+\sec(\phi_C^P+\rho)+ [\tan(\phi_C^P+\rho) / \sqrt{2\epsilon-A_{[2,3]}}  ]} \right)  \\
&\times& \left( 1-\frac{2\sec (\phi_C^P-\rho)}{1+\sec(\phi_C^P-\rho)+ [\tan(\phi_C^P-\rho) / \sqrt{2\epsilon-A_{[2,3]}} ] }\right)  \\
&=& \left( 1-\frac{2}{1+\cos(\phi_C^P+\rho)+ [\sin(\phi_C^P+\rho) / \sqrt{2\epsilon-A_{[2,3]}} ]} \right)  \\
&\times& \left( 1-\frac{2}{1+\cos(\phi_C^P-\rho)+ [\sin(\phi_C^P-\rho) / \sqrt{2\epsilon-A_{[2,3]}} ] }\right)  \\
&=& \left( 1-\frac{2}{1+\sqrt{1+\frac{1}{2\epsilon-A_{[2,3]}}} \sin(\phi_C^P+\rho+\gamma) }\right)  \\
&\times& \left( 1-\frac{2}{1+\sqrt{1+\frac{1}{2\epsilon-A_{[2,3]}}} \sin(\phi_C^P-\rho+\gamma) }\right),
\eeaa
where
\[   \sin \gamma= \frac{1}{ \sqrt{1+\frac{1}{2\epsilon-A_{[2,3]}}}}.   \]
Let's denote
\[ u=\sqrt{1+\frac{1}{2\epsilon-A_{[2,3]}}} \sin(\phi_C^P+\rho+\gamma), \quad v=\sqrt{1+\frac{1}{2\epsilon-A_{[2,3]}}} \sin(\phi_C^P-\rho+\gamma). \]
Then
\[ C=\frac{1-\sqrt{\Delta_P}}{1+\sqrt{\Delta_P}}= \frac{2(u+v)}{[\sqrt{(u+1)(v+1)}+\sqrt{(u-1)(v-1)}]^2}.\]
Since $ u>1, v>1 $, using $\sqrt{xy} \leq \frac{x+y}{2} $, one has
\be C > \frac{2(u+v)}{ [\frac{(u+1+v+1)}{2}+ \frac{(u-1+v-1)}{2} ]^2 }=\frac{2}{u+v}. \label{C} \ee
Also,
\be u+v= 2 \sqrt{1+\frac{1}{2\epsilon-A_{[2,3]}}} \sin(\phi_C^P+\gamma) \cos \rho=2\csc \gamma \sin(\phi_C^P+\gamma) \cos \rho.  \label{V} \ee
Hence  by (\ref{C}) and (\ref{V})
\[ C \cos \rho > \sin \gamma \csc (\phi_C^P+\gamma) > \sin \gamma.\]
We see that when $\epsilon \to \infty$ it has
\be  C \cos \rho    > \sin \gamma = \frac{1}{ \sqrt{1+\frac{1}{2\epsilon-A_{[2,3]}}}} > \sqrt{\frac{A_{[2,3]}}{A_{[1,4]}}}    \label{rho2}     \ee
Then by (\ref{pma}) one gets
\be  (\sqrt{A_{[1,4]}}-\sqrt{A_{[2,3]}})^2 < \hat U^P_{max} < ( A_{[1,4]}-A_{[2,3]}).\label {ep} \ee
This is like the  KP-(II) case \cite{ch, ko}. \\
\hspace*{0.3cm} On the other hand, we also consider the case $\epsilon \to 0$. From (\ref{rho}) and (\ref{V}), it follows that
\[ u+v >  2\csc \gamma \sin(\phi_C^P+\gamma) \cos\phi_C^P= 2\cos\phi_C^P( \cos\phi_C^P + \frac{\sin \phi_C^P}{\sqrt{2 \epsilon- A_{[2,3]}}}). \]
Then
\[ u+v > \frac{\sin 2 \phi_C^P}{\sqrt{2 \epsilon- A_{[2,3]}}}.\]
Therefore, it can be seen that when $\epsilon \to 0$, one has
\be  u+v >   2   \sqrt{\frac{A_{[1,4]}}{A_{[2,3]}}}                 \label{eps}            \ee
By (\ref{C})  and (\ref{eps}), we get
\[ -C >  - \sqrt{\frac{A_{[2,3]}}{A_{[1,4]}}} \]
Finally, from (\ref{pma}), it follows that
\be  (\sqrt{A_{[1,4]}}-\sqrt{A_{[2,3]}}) < \hat  U^P_{max} <  ( A_{[1,4]}+A_{[2,3]}). \label{ep2} \ee
In summary, we have, using (\ref{rho2})  and (\ref{eps}),
\\
 $  \bullet \quad  \epsilon \to \infty : \quad (\sqrt{A_{[1,4]}}- \sqrt{A_{[2,3]}})^2 < \hat U^P_{max} <  ( A_{[1,4]}-A_{[2,3]}) $   \\
 $  \bullet \quad  \epsilon \to 0 : \quad  ( A_{[1,4]}- A_{[2,3]}  ) < \hat U^P_{max} < (\sqrt{A_{[1,4]}}+ \sqrt{A_{[2,3]}}).   $
\\
\end{itemize}
\section{Concluding Remarks}
In this article, one investigates the three basic interactions of real 2-solitons solutions of the Novikov-Veselov(NV)  equation using the real Grassmannian. For the resonant $Y$-type soliton, one studies the behavior when $ y >> 0 $ and $ y << 0$. This implies that the rank condition in \cite{bc} for the KP-(II) equation would be correct for the NV equation when one considers $ y >> 0 $ and $y << 0$.  It means that for a given derangement, the soliton graph of the NV equation can be obtained from the KP-(II)'s soliton graph similarly (see (\ref{o1}) and (\ref{o2})).  As for $t \to \infty$ of the NV equation, using (\ref{T}), a similar argument in \cite{ko3} (p.25)  shows that the soliton graph can be obtained from the KP-(II)'s soliton graph. It needs further investigations. Also,  we see that the amplitudes of $O$-type and $P$-type are different from the KP-(II) equation due to the $\tan$ function. In the $O$-type soliton, the amplitude depends on $\Delta_O$ such that it possibly becomes smaller at the intersection point and in the $P$-type soliton,the amplitude depends on $\epsilon$ such that it possibly becomes bigger at the intersection point. On the other hand, for both the $O$- and $P$-type solitons, the range of this interaction angle is also founded to be limited by a critical angle which depends on the fixed amplitudes and $\epsilon$.  When  wave interaction angle between  incidence waves is outside of the prescribed range, we hope that the appropriate soliton solutions of the NV equation  can be applied as in the case of the KP-(II) equation \cite{ko5}. Finally, for the soliton interactions or the resonance of type (I) and type-(III) (\ref{ty}), they  can be considered similarly. These results will be published elsewhere.
\subsection*{Acknowledgments}
The author is grateful to Prof. Y. Kodama  for valuable discussions. He also
thanks Prof. K. Maruno for his suggestions. This work is
supported in part by the National Science Council of Taiwan under
Grant No. NSC 101-2115-M-606-001.

\begin{figure}[p]
\centering
\includegraphics[width=0.40\textwidth] {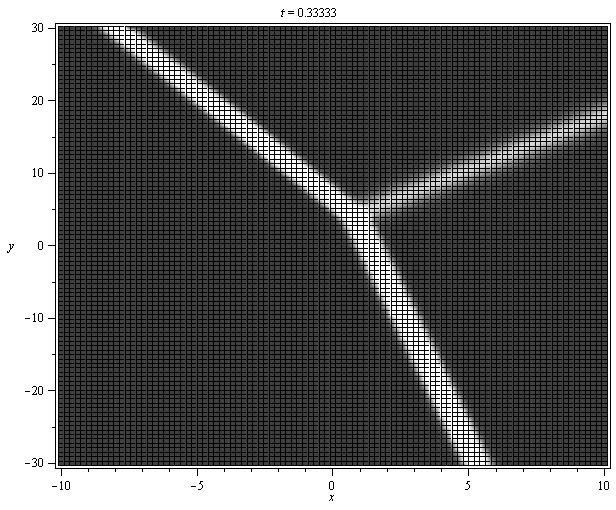}
\caption{Y-Shape, $\alpha_1=\frac{-\pi}{3}, \alpha_2=\frac{\pi}{8}, \alpha_3=\frac{\pi}{4}, \epsilon= 10, a=b=1 $  }
\end{figure}

\begin{figure}[p]
\centering
\includegraphics[width=0.40\textwidth]{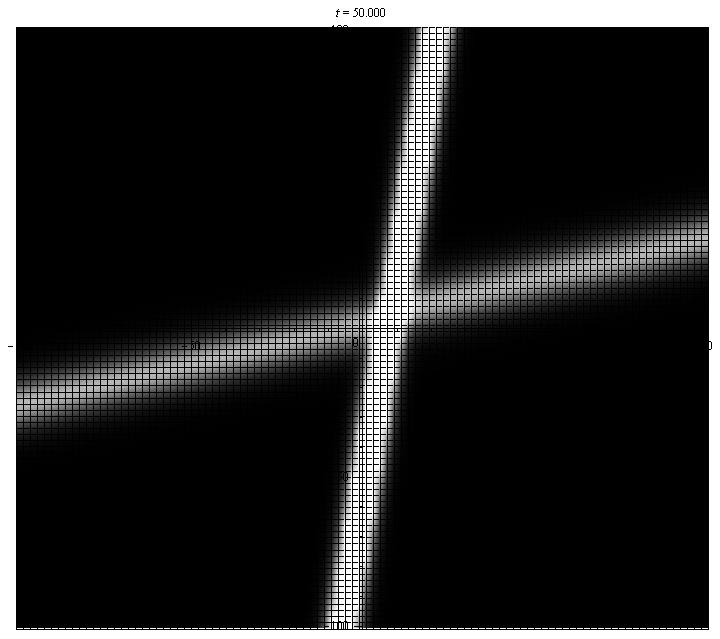}
\caption{O-type, $\alpha_1=\frac{-\pi}{6}, \alpha_2=\frac{\pi}{4}, \alpha_3=\frac{\pi}{3}, \alpha_4=\frac{\pi}{2},  \epsilon= 0.1, a=5, b=10 $  }
\end{figure}

\begin{figure}[p]
\centering
\includegraphics[width=0.40\textwidth]{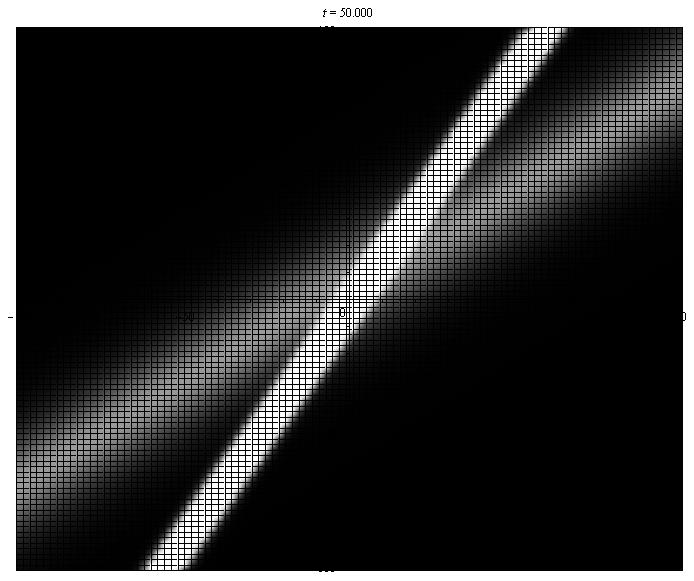}
\caption{P-type, $\alpha_1=\frac{-\pi}{6}, \alpha_2=\frac{\pi}{4}, \alpha_3=\frac{\pi}{3}, \alpha_4=\frac{\pi}{2},  \epsilon= 0.1, a=5, c=10 $  }
\end{figure}

\end{document}